%% file: HPlusWZPaper.tex
\pdfoutput=1
\newcommand*{\ATLASLATEXPATH}{}
\documentclass[cernpreprint,texlive=2011,txfonts,UKenglish]{\ATLASLATEXPATH atlasdoc}

\usepackage{\ATLASLATEXPATH atlaspackage}
\usepackage{\ATLASLATEXPATH atlasbiblatex}

\usepackage{\ATLASLATEXPATH atlascontribute}

\usepackage{\ATLASLATEXPATH atlasphysics}
\usepackage{grffile}

\graphicspath{{logos/}{figures/}}

\hypersetup{pdftitle={ATLAS draft},pdfauthor={The ATLAS Collaboration}}

\AtlasTitle{Search for a Charged Higgs Boson Produced in the Vector-boson
Fusion Mode with Decay $H^\pm \to W^\pm Z$ using  $pp$ Collisions at $\sqrt{s}=8$~TeV with the ATLAS Experiment}

\author{The ATLAS Collaboration}

\AtlasRefCode{HIGG-2014-13}

\PreprintIdNumber{CERN-PH-EP-2015-053}

\AtlasAbstract{%
 A search for a charged Higgs boson, $H^\pm$, decaying to a $W^\pm$ boson and a $Z$ boson is presented.  
The search is based on 20.3~fb$^{-1}$
  of proton-proton collision data at a center-of-mass energy of 8~TeV
recorded with the ATLAS detector at the LHC.  The $H^\pm$ boson is assumed
  to be produced via vector-boson fusion and the decays $W^\pm \to q\bar{q^\prime}$ and $Z\rightarrow
  e^+e^-/\mu^+\mu^-$ are considered.  The search is performed in a range of charged Higgs boson masses from
  $200$ to $1000$~GeV. No evidence for the production of an $H^\pm$ boson is observed.  
Upper limits of 31--1020~fb at 95\% CL are placed on the cross section for 
vector-boson fusion production of an $H^\pm$ boson times its branching fraction to $W^\pm Z$. 
The limits are compared with predictions from the Georgi-Machacek Higgs Triplet Model.
}

\AtlasCoverSupportingNote{\textsl{•}}{https://cds.cern.ch/record/1957180}

\AtlasJournal{PRL}

 \AtlasCoverAnalysisTeam{
 C.~Gwilliam,
 A.~Mehta
 }

\AtlasCoverEdBoardMember{Marcel Vos~(chair),Viviana Cavaliere,Peter Maettig,Daniel Whiteson}

\nolinenumbers

\newcommand\mjj{\ensuremath{m_{jj}}}
\newcommand\mll{\ensuremath{m_{\ell \ell}}}
\newcommand\mlljj{\ensuremath{m_{\ell \ell jj}}}

\newcommand\dphill{\ensuremath{\Delta\phi_{\ell \ell}}}

\begin{document}
\label{sec:intro}
After the discovery of a Higgs boson at the LHC in 2012~\cite{paper2012ichep,CMSpaper}, an important
question now is whether the newly discovered particle is part of an extended
scalar sector. The discovery of additional scalar or pseudoscalar bosons would provide
spectacular evidence that this is the case.

A charged Higgs boson, $H^\pm$, appears in many models with an
extended scalar sector such as the Two Higgs Doublet
Model, where a second Higgs doublet is introduced~\cite{Branco:2011iw}, and the Higgs Triplet
Models~\cite{georgimachacek, HTriplet}, where a triplet is added to the Higgs doublet of the Standard Model (SM). 
While $H^{\pm} \to \tau^\pm\nu, cs, tb$ decays 
dominate in the Two Higgs Doublet Model
at tree level, $H^{\pm} \to W^\pm
Z$ decays are allowed at loop level~\cite{THDM} and are predicted
 at tree level in Higgs Triplet Models. 
In the search presented in this Letter,
 the $H^\pm$ boson is assumed to couple to $W^\pm$ and $Z$ bosons. In this case it is
 produced via vector-boson
fusion  (VBF), $W^\pm Z \to H^\pm$, at the LHC and decays to $W^\pm Z$.  The search is performed in the channel
with subsequent decays of $W^\pm \to 
q\bar{q^\prime}$ and $Z \to \ell^+ \ell^-$, over the $H^\pm$ mass
 range $200< m_{H^\pm}< 1000$~GeV. The data are compared with the Georgi-Machacek  Higgs Triplet Model (GMHTM)~\cite{georgimachacek}.

Searches at LEP have looked
for pair-produced charged Higgs bosons~\cite{Abbiendi:2013hk}.  Previous searches  at
the Tevatron and LHC have looked for a
charged Higgs boson produced in top quark decays or in associated production with a top quark~\cite{Abazov:2009aa,  Chatrchyan:2012vca, Aaltonen:2014hua,Aad:2014kga}.
 Searches for a $W^\pm Z$
resonance have also been performed in non-Higgs-specific models~\cite{Chatrchyan:2012kk,
Chatrchyan:2012rva, Aad:2012vs, Aad:2013wxa, Khachatryan:2014gha,
Aad:2014pha,Khachatryan:2014xja}, but this search is the first to
look  specifically for the VBF production mechanism.

\label{Sec:samples}
The data used in this search were recorded with the ATLAS detector in proton-proton collisions at a
center-of-mass energy of $\sqrt{s}=8$~TeV.   In the ATLAS coordinate system, the polar angle $\theta$ is measured
with respect to the LHC beam-line and  the azimuthal angle $\phi$ is measured in the plane transverse
to the beam-line. Pseudorapidity  is defined as $\eta = -\ln \tan (\theta /2)$.

The ATLAS detector is described in detail
elsewhere~\cite{Aad:2008zzm}. It consists of an inner tracking
detector covering the range $|\eta|<2.5$, surrounded by a
superconducting solenoid providing a 2~T magnetic field,
electromagnetic and hadronic calorimeters ($|\eta|<4.9$) and an
external muon spectrometer ($|\eta|<2.7$), consisting of three air-core toroidal magnets, 
interspersed with high precision tracking chambers.
  The integrated luminosity
of the data sample, considering only data-taking periods where all
relevant detector subsystems were operational, is 20.3~$\pm$~0.6~${\rm
fb}^{-1}$~\cite{lumi}.  The data were collected using a combination of
single-electron, single-muon, electron-electron and muon-muon
triggers.  The higher level $\pt$ thresholds for both electrons and muon triggers
 are 24~GeV for the single-lepton
triggers and 13~GeV for the dilepton triggers.

Monte Carlo simulated events are used to estimate distributions of expected
signal and background events.
Signal events are generated for a narrow-width $H^\pm$ boson produced via VBF with
MADGRAPH5~\cite{Alwall:2011uj} using CTEQ6L1~\cite{Pumplin:2002vw} parton distribution functions (PDFs). The parton showering is performed with 
PYTHIA8~\cite{Pythia,Pythia8}.
The dominant background is the production of $Z$ bosons in association with jets, which is
 simulated
with SHERPA~\cite{SHERPA} using CT10 PDFs~\cite{ct10}. Top quark pair, single top quark and diboson production 
are simulated with POWHEG~\cite{Powheg1,Powheg2,Powheg3} using  CTEQ6L1 PDFs. 

All simulated samples are passed through the ATLAS GEANT4-based 
detector simulation~\cite{AGO-0301,ATLASSIM}.
The simulated events are  overlaid with additional minimum-bias events
to account for the effect of multiple $pp$ interactions (pile-up) occurring in the same 
and neighboring bunch crossings~\cite{Aad:2014bia}.

\label{sec:objects}

Electrons are identified for $|\eta| < 2.47$ and $\pt > 7$~GeV from energy clusters in the
electromagnetic calorimeter that are matched to tracks in the inner detector \cite{egammapaper}.
Quality requirements on the  calorimeter clusters and tracks are applied to
reduce contamination from jets. The jet background is further reduced by applying isolation requirements
that are based on tracking information within a cone around the electron candidate~\cite{Aad:2014xzb}.

Muons are reconstructed in the muon spectrometer in the range $|\eta|< 2.7$ and $\pt >
 7$~GeV~\cite{Aad:2014rra}.  For $|\eta|< 2.5$ the muon spectrometer track must be matched with a
 track in the inner detector and information from both is used to reconstruct the momentum. The muon
 candidates are required to pass isolation requirements similar to those for electrons.

Jets are reconstructed using the anti-$k_t$ algorithm~\cite{Cacciari:2008gp} with size
parameter \mbox{$R = 0.4$} and are restricted to  $|\eta| < 4.5$ in order to fully contain 
each of the jets in the calorimeters. 
 Those jets with $|\eta| < 2.5$, 
where there is good tracking coverage, are called central jets and are 
required to have  $\pt>20$~GeV. 
Those jets with $2.5<|\eta| < 4.5$ are required to have $\pt>30$~GeV. 
Low-$\pt$ central  jets from pile-up are suppressed  with the following
requirement: 
for jets with $|\eta| < 2.4$ and $\pt < 50$~GeV, tracks associated with the primary vertex must 
contribute over 50\% to the scalar sum of the $\pt$ of all the tracks associated with the jet.
Jets originating 
from $b$-quark fragmentation are 
selected using a multivariate tagging algorithm ($b$-tagging)~\cite{ATLAS-CONF-2014-004}.  The
$b$-tagging algorithm is only applied to central jets and its operating point is chosen such that the efficiency to select $b$-quark jets is approximately 70\%.

The magnitude of the missing transverse momentum ($\met$) is computed using fully calibrated electrons, muons, jets and calorimeter clusters not associated to other physics objects~\cite{ATLAS-CONF-2013-082}.

\label{sec:selection}

The $Z \rightarrow \ell^+ \ell^-$ decay is reconstructed  from two electrons or 
two muons. In events with
muons, where there is a very low charge misidentification
probability, the leptons must be oppositely charged.  In order to match  the single-lepton trigger
threshold and reduce the
multijet background, tighter requirements are placed on one of the leptons.  These requirements are that the lepton has $\pt>25$~GeV and, if it is a muon, is restricted to 
$|\eta|<2.5$.  The mass of the $Z$ boson candidate is reconstructed from the
two leptons and must satisfy $83<\mll<99$~GeV.

The VBF process generally contains two reconstructed jets,
referred to as tag jets, with high $|\eta|$ in opposite directions. 
 The tag-jet selection begins by requiring two non-$b$-tagged jets in 
 opposite hemispheres.  If more than one such pair is found,  the
one with the highest invariant mass is selected.   The
tag-jet pair must have an invariant mass greater than $500$~GeV and $|\Delta \eta|>4$.

Once a tag-jet pair has been identified, the  $W^\pm \to q\bar{q^\prime}$ decay
 is reconstructed from the two highest-$\pt$ remaining central
 jets. These jets are referred to as signal jets.  In order to reduce the 
$Z+$jets background, at least one signal jet is required to have $\pt>45\gev$. A cut on the dijet invariant mass of $60 < \mjj\ < 95$~GeV, consistent with the $W^\pm$ mass, is made.

Background from top quark production is reduced by rejecting events with two
or more $b$-tagged jets and requiring the $\met$ significance $\met/\sqrt{H_{\text{T}}}<6$~${\rm GeV}^{0.5}$,
where $H_{\text{T}}$ is the scalar sum of the transverse momenta of all jets and leptons in the event.

The invariant mass of the two leptons and two signal jets $\mlljj$ is used to reconstruct the charged Higgs boson mass,
$m_{H^\pm}$. The resolution is improved by using the $W^\pm$ mass~\cite{Agashe:2014kda}, $m_W=80.4$~GeV, as a constraint by
scaling the energy of each jet by $m_W/m_{jj}$. The resulting experimental resolution on $\mlljj$, determined from simulation, is on average $2.4\%$ and is approximately independent of $\mlljj$ over the range of the analysis.

In order to reduce the $Z+$jets background, cuts are imposed on the transverse momentum and
azimuthal angular separation of the lepton pair: $\pt^{\ell \ell} > {\rm min}[ 0.46\, \mlljj-54 \, {\rm GeV}, 275 \, {\rm GeV} ]$ and $\dphill < 1+(270 \, {\rm GeV} /\mlljj)^{3.5}$, and also on the transverse momentum
of the signal jets $\pt^j > 0.1\mlljj$. These cuts depend on the reconstructed charged Higgs boson mass
since the decay products of the $W^\pm$ and $Z$ bosons in signal events tend to be at higher transverse momentum and more collimated
as $\mlljj$  increases. The cut values were chosen  which 
 maximized the  sensitivity of the analysis.

After all cuts, a total of 506 data events are selected.  The efficiency of the
 signal selection is 5\% for a narrow width $H^\pm$ at $m_{H^\pm}=200$~GeV, rising to a maximum
 of 9\% at $m_{H^\pm}=600$~GeV. The efficiency rises with increasing  $m_{H^\pm}$ 
 because the $W$ and $Z$ bosons receive a higher transverse boost, so less events fail the minimum transverse momentum requirements on the leptons and jets.  The efficiency falls to 2\% at 1 TeV since the products from the $W$ decay begin to overlap to form a single jet.

\label{sec:backgrounds}
The dominant background after all cuts is $Z$+jets. Other smaller backgrounds that are taken into account are top pair production, single top production, diboson production and multijet background.
The shapes of all backgrounds are determined from simulation, 
apart from the multijet background, which is estimated from data.  

In the electron channel, the multijet background is determined by
selecting a sample of events with a reversed isolation requirement on
the electron. This is normalized by performing a fit to the $m_{\ell\ell}$
distribution of the data, with the normalizations of the multijet
background and of a template made from the sum of the remaining
backgrounds left as free parameters. The systematic error on the
multijet normalization is evaluated to be 50\%. It is determined by
looking at the difference in scale factors obtained for a sample of
events with either two or three central jets. The multijet background
in the muon channel is found to be negligible.

The normalization  of the $Z$+jets background is determined from 
the signal region  by leaving it as a free parameter
of the profile likelihood fit as discussed below.
The contribution to the expected number of events  in the signal 
region due to top quark production is constrained by comparing the 
observed and expected yields in a control region enriched in top quark 
pairs. This control region is defined by selecting events with the same cuts as
those for the signal region but with an electron and a muon, rather 
than two same-flavor leptons, and two $b$-tagged jets with 
$50<m_{jj}<180$~GeV replacing the central-jet requirements. 
A total of 261 data events are selected.
The other backgrounds, diboson and single top quark production, are normalized according to the theory cross section calculated at 
next-to-leading order as listed in Ref.~\cite{Aad:2014xzb}.

\label{sec:systs}

The largest systematic uncertainties arise from the signal acceptance modeling, the normalization
and modeling of the $Z$+jets/top backgrounds and from the jet energy scale.
 
The jet energy scale systematic uncertainty arises from several sources including uncertainties from
the in-situ calibration, pile-up-dependent corrections and the jet flavor
composition~\cite{Aad:2014bia}. A systematic error on the jet energy resolution is also included.
These uncertainties are propagated to the $\met$, which also has a contribution from hadronic energy
that is not included in jets~\cite{ATLAS-CONF-2013-082}.  The uncertainty in the pile-up is
accounted for by varying the cut against pile-up jets and varying the assumed number of pile-up
interactions in the simulated events.  The $b$-tagging efficiency uncertainty is dependent on jet
$\pt$ and comes mainly from the uncertainty on the measurement of the efficiency in top quark pair
events~\cite{ATLAS-CONF-2014-004}.
Other experimental systematic uncertainties that are included arise from the lepton energy scale,
lepton identification efficiency and the uncertainty on the multijet background prediction.

In addition to the experimental systematic uncertainties, modeling systematic uncertainties are
included to account for possible differences between the data and the simulation model that is used
for each background process, following closely the procedure described in Ref.~\cite{Aad:2014xzb}.
The $Z$+jets background includes uncertainties on the relative fractions of the different flavor
components, the shape of distributions of $m_{jj}$, the azimuthal separation of the central jet pair
and the transverse momentum of the lepton pair.  For top quark pair production, uncertainties on the
top quark transverse momentum and $m_{jj}$ distributions are included.  Uncertainties on the ratio
of the numbers of events containing two and three reconstructed signal jets are also included for
each background.

Modeling uncertainties on the signal acceptance are taken into account by varying the factorization
and normalization scale up and down by a factor of two, varying the amount of initial- and
final-state radiation and comparing the default CTEQ6L1 PDFs to MSTW2008lo68cl~\cite{Martin:2009iq}
and NNPDF21\_lo\_as\_0119\_100~\cite{Ball:2011mu} PDFs.  The combined signal acceptance uncertainty
is $\sim 10\%$ and approximately constant with $m_{H^\pm}$.

The contribution of the various sources of uncertainty for an example
production scenario is given in Table~\ref{tab:systs}.

\begin{table}[htb]
\centering
\begin{tabular}{l c}
\hline\hline
Source of uncertainty & Impact (\%) \\
\hline
Total & 20\\
Statistical & 14\\
Systematic & 14\\

\hline
\multicolumn{2}{c}{Experimental Uncertainties}\\
\hline

Jets$+E_T^{\rm miss}$ & 8.8\\
Luminosity &  2.8\\
Leptons & 0.9\\
$b$-tagging & 0.2\\

\hline
\multicolumn{2}{c}{Theoretical and Modeling Uncertainties}\\
\hline

Signal & 10\\
Top & 3.8\\
$Z$+jets & 3.7\\
Multijet & 0.1\\

\hline\hline
\end{tabular}
\caption{\label{tab:systs} The percentage impact of the various
  sources of uncertainty on the expected  $H^{\pm}$ production cross section for  
$m_{H^\pm}=400$~GeV and
   $\sigma\times BR(H^{\pm} \to W^{\pm}Z) = 1$~pb.}
\end{table}

\label{sec:results}
\begin{figure}[h]
  \centering
\includegraphics[width=0.65\textwidth]{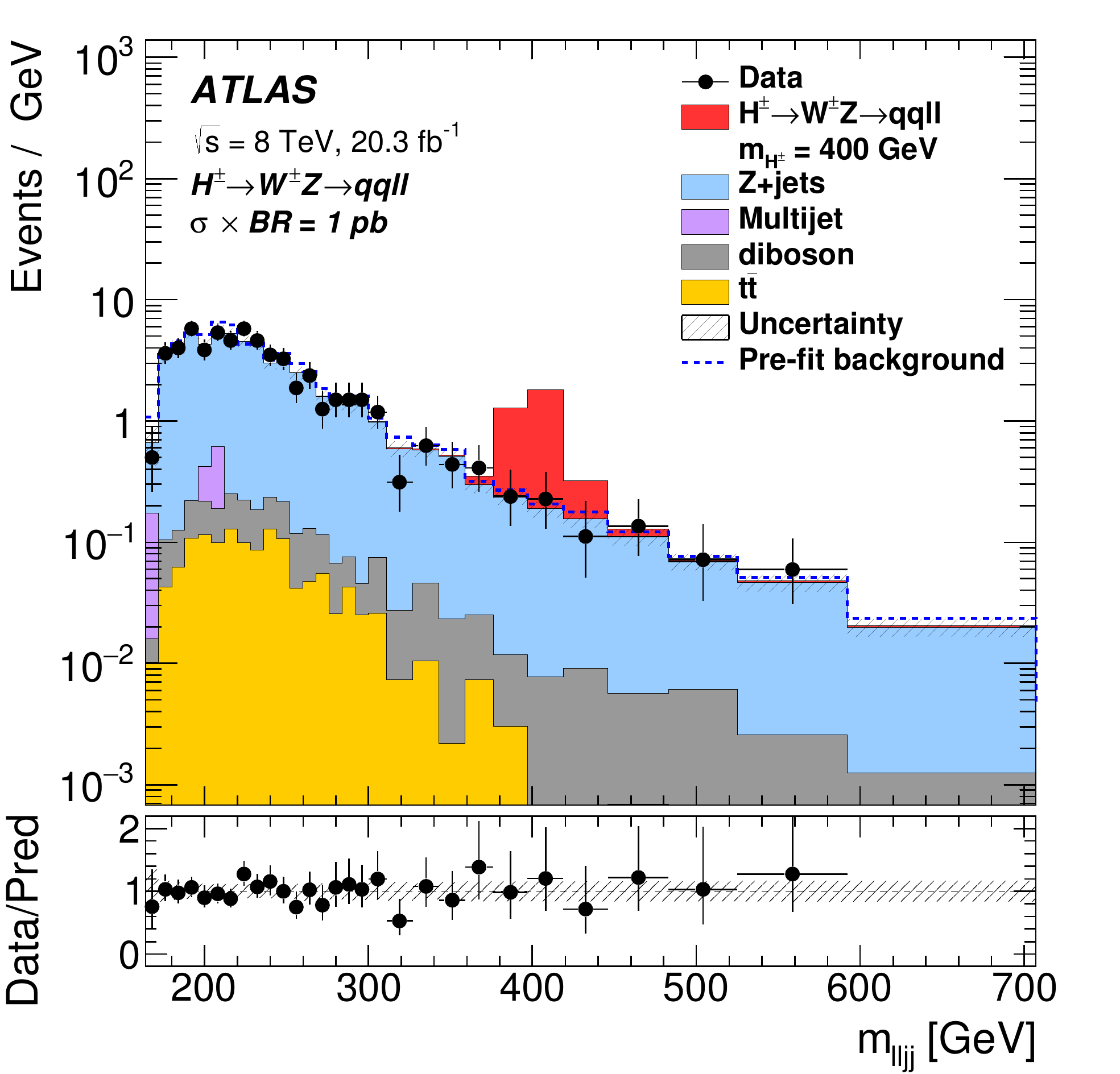}
\caption{\label{fig:mlljj} The $m_{\ell\ell jj}$ distribution for data and the expected SM background. 
  The hashed band indicates the post-fit systematic uncertainty.  Included in the plot is an example
signal sample with $m_{H^\pm}=400$~GeV, which has been plotted with a cross section times
branching fraction,  $\sigma\times$BR($H^\pm\to W^\pm Z$) $=1$ pb for illustration. No data events are observed for values of  $m_{\ell\ell jj}>600$~GeV.}
\end{figure}

The data are compared with the  SM expectation in Fig.~\ref{fig:mlljj}.  The expectation is
determined with a profile likelihood fit~\cite{CLs_2002} using the modified frequentist method, also
known as ${\rm CL_s}$~\cite{Cowan2011}. The fit is performed on the $\mlljj$ mass distribution in the signal region and the
number of events in the top quark control region.  No significant excess of events is observed in
the data compared with the SM expectation.

\begin{figure}[h]
  \centering
\includegraphics[width=0.65\textwidth]{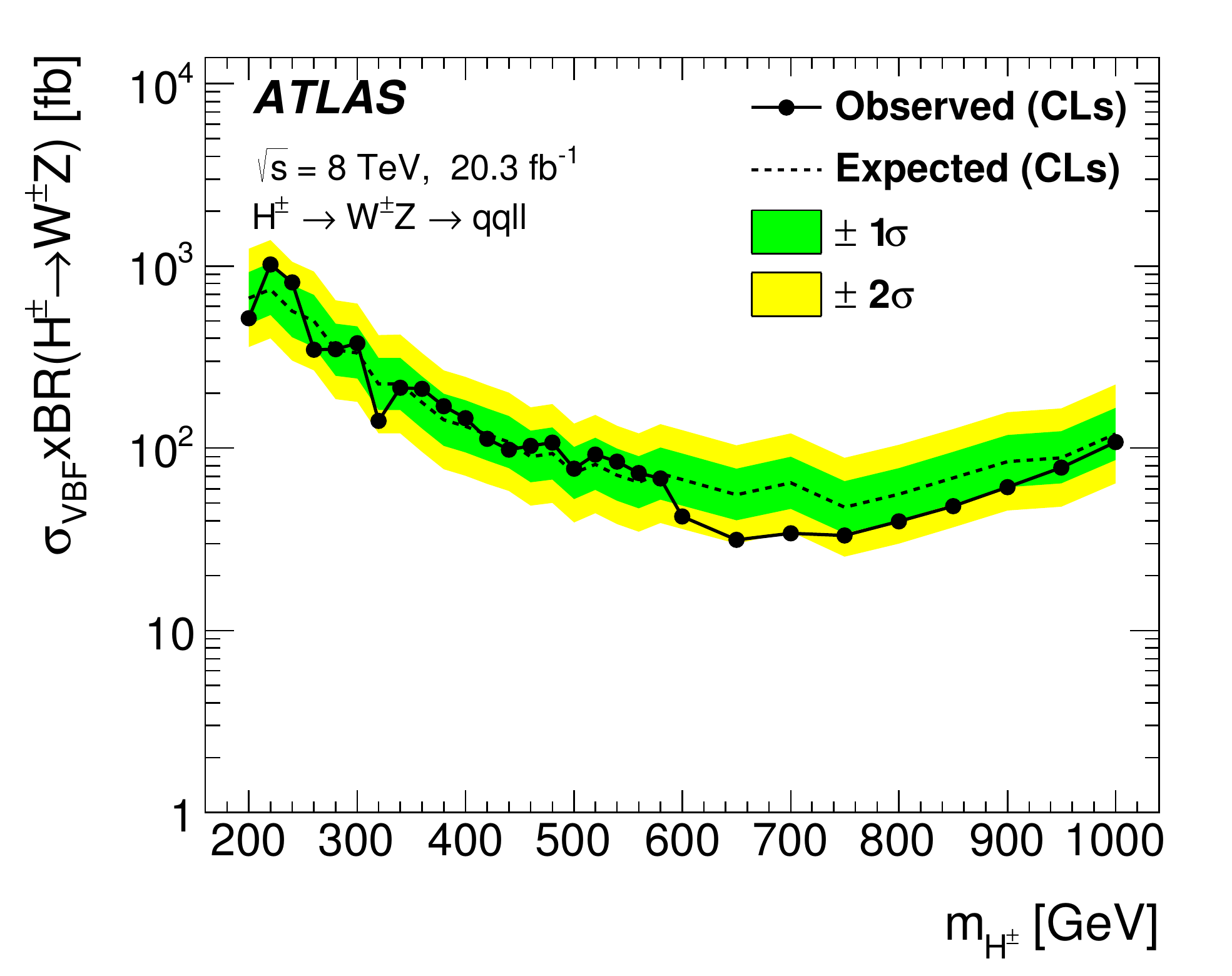}
  \caption{Exclusion limits in fb at the 95\% CL for the  vector-boson fusion production cross section of a  $H^\pm$ boson times its branching fraction to $W^\pm Z$  assuming the signal has a  narrow intrinsic  width. 
Also included on the plot are the median, $\pm 1 \sigma$  and  $\pm 2 \sigma$ values within which the limit is expected to lie in the absence of a signal.}
\label{fig:limit}
\end{figure}

 Figure~\ref{fig:limit} shows the  exclusion limits at the 95\% confidence level (CL) on the VBF production
 cross section times the branching fraction BR($H^\pm\to W^\pm Z$) as a function of $m_{H^\pm}$, assuming the signal has a small intrinsic width, i.e. much
smaller than the experimental resolution.  The
 observed limits range from 31~fb at $m_{H^\pm}=650$~GeV to 1020~fb at $m_{H^\pm}=220$~GeV; the
 corresponding expected limits are 55~fb and 719~fb, respectively. The observed limits are
  better for masses less than around 800 GeV than those obtained the inclusive ATLAS $WZ$ 
resonance search, presented in Ref.\cite{Aad:2014pha},  by up to a factor of six.

\begin{figure}[h]
  \centering
\includegraphics[width=0.65\textwidth]{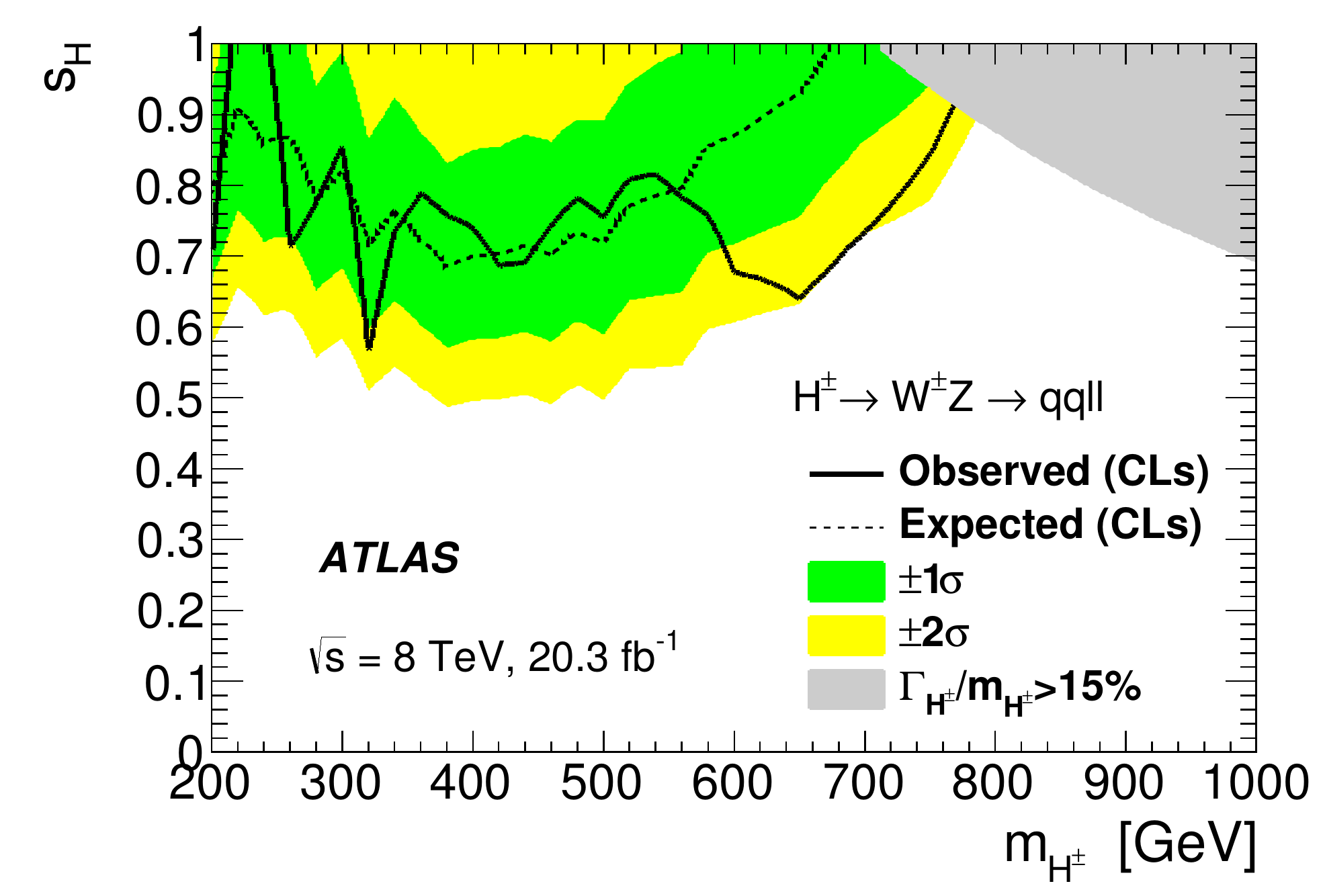}
  \caption{Exclusion limits  at the 95\% CL for $s_H$  versus $m_{H^\pm}$ in the 
 Georgi-Machacek Higgs Triplet Model. Also included on the plot are the median, $\pm 1 \sigma$  and  $\pm 2 \sigma$ values within which the limit is expected to lie in the absence of a signal. }
\label{fig:limitGMHTM}
\end{figure}

The exclusion limits are  compared with  predicted charged Higgs boson
production cross sections in the GMHTM~\cite{Bolzoni:2011cu},
calculated at  $\sqrt{s}=8$~TeV~\cite{HiggsTriplet8TeV}. 
In this model,  the quantity $s_H^2$ is a free parameter that 
 represents the fraction of the square of the  gauge boson masses $m_W^2$ and $m_Z^2$ that is generated by the vacuum expectation  value of the triplet, 
while $1-s_H^2$ represents the fraction generated by the  SM Higgs doublet. 
The production cross section and $H^\pm$ width are proportional to $s_H^2$. 
The branching fraction of $H^\pm \to
 W^\pm Z$ is expected to be very high above the $W^\pm Z$ threshold, so for simplicity it is set to one.
For this  comparison  a nonzero intrinsic $H^\pm$ width must be
 taken into account. 
 This is done by smearing the signal $\mlljj$
 distributions with a relativistic Breit-Wigner distribution with a width as
 calculated in Ref.~\cite{HiggsTriplet8TeV}.  The fractional width $\Gamma_{H^\pm} /m_{H^\pm}$  for $s_H=1$ 
  increases from $0.2\% $ at $m_{H^\pm}=200$~GeV to $31\%$ at
 $m_{H^\pm}=1000$~GeV. Comparisons with the GMHTM are only shown
for $\Gamma_{H^\pm}/m_{H^\pm}< 0.15$, since  higher values
may violate perturbative  unitarity of the $W^\pm Z \to W^\pm Z$ scattering 
amplitudes.
 As shown in  Fig.~\ref{fig:limitGMHTM}, the data exclude a charged
 Higgs boson over the range $240 <m_{H^\pm} <700$~GeV for $s_H=1$, 
with weaker limits for smaller values of $s_H$.

 \FloatBarrier 
\label{sec:summary}
In conclusion, data recorded by the ATLAS experiment at the LHC, corresponding to an integrated luminosity
of 20.3~fb$^{-1}$ at a center-of-mass energy of 8~TeV,
have been used to search for a charged Higgs boson, produced via 
vector-boson fusion and decaying to $W^\pm Z$, over the charged Higgs boson mass range
200--1000~GeV. This is the first search for this process.
No deviation from the SM background prediction is observed.
Upper limits at the 95\% confidence level on the
cross section of a VBF-produced $H^\pm$ boson times its branching fraction to $W^\pm Z$
are set between 31 and 1020~fb for a narrow $W^\pm Z$ resonance.
The data exclude a charged Higgs boson in the range 
$240 <m_{H^\pm} <700$~GeV within the Georgi-Machacek  Higgs Triplet Model
with  parameter $s_{H}=1$ and 100\% branching fraction of $H^\pm \to
 W^\pm Z$.

We thank CERN for the very successful operation of the LHC, as well as the
support staff from our institutions without whom ATLAS could not be
operated efficiently.

We acknowledge the support of ANPCyT, Argentina; YerPhI, Armenia; ARC,
Australia; BMWFW and FWF, Austria; ANAS, Azerbaijan; SSTC, Belarus; CNPq and FAPESP,
Brazil; NSERC, NRC and CFI, Canada; CERN; CONICYT, Chile; CAS, MOST and NSFC,
China; COLCIENCIAS, Colombia; MSMT CR, MPO CR and VSC CR, Czech Republic;
DNRF, DNSRC and Lundbeck Foundation, Denmark; EPLANET, ERC and NSRF, European Union;
IN2P3-CNRS, CEA-DSM/IRFU, France; GNSF, Georgia; BMBF, DFG, HGF, MPG and AvH
Foundation, Germany; GSRT and NSRF, Greece; RGC, Hong Kong SAR, China; ISF, MINERVA, GIF, I-CORE and Benoziyo Center, Israel; INFN, Italy; MEXT and JSPS, Japan; CNRST, Morocco; FOM and NWO, Netherlands; BRF and RCN, Norway; MNiSW and NCN, Poland; GRICES and FCT, Portugal; MNE/IFA, Romania; MES of Russia and ROSATOM, Russian Federation; JINR; MSTD,
Serbia; MSSR, Slovakia; ARRS and MIZ\v{S}, Slovenia; DST/NRF, South Africa;
MINECO, Spain; SRC and Wallenberg Foundation, Sweden; SER, SNSF and Cantons of
Bern and Geneva, Switzerland; NSC, Taiwan; TAEK, Turkey; STFC, the Royal
Society and Leverhulme Trust, United Kingdom; DOE and NSF, United States of
America.

The crucial computing support from all WLCG partners is acknowledged
gratefully, in particular from CERN and the ATLAS Tier-1 facilities at
TRIUMF (Canada), NDGF (Denmark, Norway, Sweden), CC-IN2P3 (France),
KIT/GridKA (Germany), INFN-CNAF (Italy), NL-T1 (Netherlands), PIC (Spain),
ASGC (Taiwan), RAL (UK) and BNL (USA) and in the Tier-2 facilities
worldwide.

\printbibliography
\clearpage

\clearpage
\input{atlas_authlist}
\end{document}

%% file: atlas_authlist.tex
\begin{flushleft}
{\Large The ATLAS Collaboration}

\bigskip

G.~Aad$^{\rm 85}$,
B.~Abbott$^{\rm 113}$,
J.~Abdallah$^{\rm 151}$,
O.~Abdinov$^{\rm 11}$,
R.~Aben$^{\rm 107}$,
M.~Abolins$^{\rm 90}$,
O.S.~AbouZeid$^{\rm 158}$,
H.~Abramowicz$^{\rm 153}$,
H.~Abreu$^{\rm 152}$,
R.~Abreu$^{\rm 30}$
% ...
\bigskip
\\
% ...
$^{11}$ Institute of Physics, Azerbaijan Academy of Sciences, Baku, Azerbaijan\\
% ...
$^{30}$ CERN, Geneva, Switzerland\\
% ...
$^{85}$ CPPM, Aix-Marseille Universit{\'e} and CNRS/IN2P3, Marseille, France\\
% ...
$^{90}$ Department of Physics and Astronomy, Michigan State University, East Lansing MI, United States of America\\
% ...
$^{107}$ Nikhef National Institute for Subatomic Physics and University of Amsterdam, Amsterdam, Netherlands\\
% ...
$^{113}$ Homer L. Dodge Department of Physics and Astronomy, University of Oklahoma, Norman OK, United States of America\\
% ...
$^{151}$ Institute of Physics, Academia Sinica, Taipei, Taiwan\\
$^{152}$ Department of Physics, Technion: Israel Institute of Technology, Haifa, Israel\\
$^{153}$ Raymond and Beverly Sackler School of Physics and Astronomy, Tel Aviv University, Tel Aviv, Israel\\
% ...
$^{158}$ Department of Physics, University of Toronto, Toronto ON, Canada\\
% ...
\end{flushleft}
